\documentstyle[preprint,aps]{revtex}

\newcommand{\be}{\begin{equation}}
\newcommand{\ee}{\end{equation}}
\newcommand{\bea}{\begin{eqnarray}}
\newcommand{\eea}{\end{eqnarray}}
\newcommand{\ba}{\begin{array}}
\newcommand{\ea}{\end{array}}

\newcommand{\ler}{\stackrel{\scriptstyle <}{\scriptstyle\sim}}
\newcommand{\ger}{\stackrel{\scriptstyle >}{\scriptstyle\sim}}

\begin{document}
\preprint{
KAIST-TH 3/97 
}


\title{\bf Axions and the Strong CP Problem in 
$M$-theory}
\author{Kiwoon Choi
\thanks{kchoi@chep6.kaist.ac.kr}}
\address{
Department of Physics,
Korea Advanced Institute of
Science and Technology,
\\
Taejon 305-701, Korea}

\maketitle

\def\be{\begin{equation}}
\def\ee{\end{equation}}
\def\ap{\approx}

\begin{abstract}
We examine the possibility that 
the strong CP problem is solved by  
string-theoretic axions  
in strong-coupling limit of the $E_8\times E^{\prime}_8$
heterotic string theory ($M$-theory).
We first discuss some generic features of gauge kinetic functions
in compactified $M$-theory, and examine in detail
the axion potential induced
by the explicit breakings other than the QCD anomaly of the non-linear
$U(1)_{PQ}$ symmetries of string-theoretic axions.
It is argued based on supersymmetry and
discrete gauge symmetries that if the compactification
radius is large enough, there can be  a
$U(1)_{PQ}$-symmetry whose breaking other than the QCD anomaly, 
whatever its microscopic origin is,
is suppressed enough for 
the axion mechanism to work.
Phenomenological viability of such a large radius crucially
depends upon the quantized coefficients in gauge kinetic functions.
We note that 
the large radius required for the axion mechanism is viable
only in a  limited class of models.
For instance, for compactifications on
a smooth Calabi-Yau manifold
with a vanishing $E^{\prime}_8$ field strength,
it is viable  only when the quantized flux
of the antisymmetric tensor field in $M$-theory has a minimal
nonzero value.
It is also stressed that this large compactification radius
allows  the QCD axion in $M$-theory to be cosmologically viable
in the presence of a late time entropy production.

\end{abstract}


\narrowtext

\section{Introduction and Summary}

It has long been known that 
$E_8\times E^{\prime}_8$ heterotic
string theory compactified on  a {\it large} internal six manifold
gives rise to an approximate $U(1)_{PQ}$ symmetry which may solve
the strong CP problem by means of the axion mechanism \cite{witten,choi}. 
Compactified string models always contain the model-independent axion
which corresponds to the massless mode of the antisymmetric tensor
field in flat spacetime direction.
If the internal manifold admits harmonic two forms,
which is always the case for the compactification preserving $N=1$
four-dimensional supersymmetry,
the model contains additional model-dependent axions.
As is well known, in order for the strong CP problem
to be solved by the axion mechanism, one needs 
a global $U(1)_{PQ}$ symmetry whose {\it explicit}
breaking is almost entirely given by the QCD anomaly $(F\tilde{F})_{QCD}$ 
\cite{pq}. 
The $U(1)_{PQ}$-breaking 
other than the QCD anomaly must be so tiny that its contribution
to the axion potential satisfies 
\be
\delta V_{\rm axion}\ler 10^{-9} 
f_{\pi}^2m_{\pi}^2\approx 10^{-13} \, {\rm GeV}^4.
\ee

Although string-theoretic axions 
have the desired coupling to the QCD anomaly,
they have  potentially harmful additional couplings which would make  
these axions useless for the strong CP problem.
For instance, they generically couple
to the hidden sector anomaly $(F\tilde{F})_{\rm hid}$
\cite{choi}
and also have non-derivative couplings
induced by string world-sheet instantons \cite{dine}.
As we will discuss later in detail,
$U(1)_{PQ}$-breaking by the hidden sector anomaly 
generically leads to an axion potential much bigger than
$10^{-9}f_{\pi}^2 m_{\pi}^2$, and then
it must be avoided.
In fact, in many compactification models,
there exists a linear combination
of string-theoretic axions which does not couple to the hidden sector
anomaly, while keeping the coupling to the QCD anomaly \cite{choi}.
However such a combination still receives a high energy potential
$\delta V_{\rm axion}\ap e^{-2\pi T} m_{3/2}^2M_P^2$  
from string world-sheet instantons (or equivalently
membrane-instantons
in $M$-theory) where $T$ corresponds to
the radius-squared of the compact six manifold in the unit of heterotic
string tension \cite{dine}.

If the compactification radius is large enough
so that ${\rm Re}(T)\ger 20$,  world-sheet instanton
effects would be small enough to satisfy the bound
(1).
It has been known that such a large radius is {\it not}
likely to be allowed within a weakly coupled heterotic
string theory \cite{kaplunov}, and thus its possibility has  not been taken
seriously. 
Recently Banks and Dine \cite{bank} argued that the large 
compactification radius
giving ${\rm Re}(T)\ger 20$
can be realized  in strong-coupling limit
of the $E_8\times E_8$ heterotic string theory, i.e.
in $M$-theory limit \cite{horava}, opening the possibility
for the axion solution to the strong CP problem 
in the context of $M$-theory.
However it has been  pointed out subsequently \cite{kaplu}
that there is a certain limitation
on the compactification radius even in $M$-theory,
which would require a more careful analysis
on this problem.

The purpose of this paper is to carefully examine under what conditions 
the strong CP problem can be solved by string-theoretic axions
in $M$-theory.
In the next section, we set up the notations for later
analysis by summarizing  the known facts on the
axions,  couplings and scales in heterotic string
and $M$-theories.
All couplings and scales are expressed 
in terms of the 
dilaton and K\"{a}hler modulus  superfields, $S$ and $T$, which makes
it convenient to interpret them in the context of four-dimensional
effective supergravity model.
In sect. III, we discuss some model-independent features of holomorphic
gauge kinetic functions  in the effective supergravity 
of heterotic string 
and $M$-theories.
Particular attention is paid for the quantized coefficients
of the K\"{a}hler moduli superfields $T_I$ which would  determine  
the phenomenological viability of the axion mechanism
in $M$-theory.

The non-linear $U(1)_{PQ}$ symmetries associated with string-theoretic
axions are {\it explicitly} broken not only by the desired QCD anomaly,
but also  by the potentially harmful 
hidden sector gauge anomaly and/or more microscopic 
stringy ($M$-theoretic)  effects, e.g. the world-sheet
(membrane) instantons.
In sect. IV,   we analyze in detail the high energy potential of the
model-independent
axion due to the hidden sector anomaly.
The resulting axion potential  
does {\it not} satisfy the bound (1)  {\it unless} the model is 
carefully tuned 
to forbid all dangerous non-renormalizable
operators including not only the hidden sector fields
but also the observable sector fields. 
Our study suggests that it is {\it unlikely} that 
the strong CP problem is 
solved by the model-independent axion alone, or at least 
implementing such a scenario appears to be much more nontrivial than
what has appeared  in the previous works \cite{bank1,wu}.

In sect. V, we consider  a $U(1)_{PQ}$-symmetry  
associated with a linear
combination of the model-independent axion and 
the model-dependent K\"{a}hler axions, which is designed to 
avoid the hidden sector anomaly.
An issue that should be taken account of when one considers
the axion solution to the strong CP problem is the possibility
of $U(1)_{PQ}$ breaking by generic quantum gravity effects \cite{holman}.
In this regard, it is desirable that some gauge symmetries
protect $U(1)_{PQ}$ from potentially dangerous microscopic 
stringy ($M$-theoretic) effects. 
We  argue  that supersymmetry and the discrete
gauge symmetries highly constrain the explicit breaking
of  our $U(1)_{PQ}$, and as a result 
if the compactification
radius is large enough to yield ${\rm Re}(T)\ger 20$,
potentially harmful breaking
of $U(1)_{PQ}$ other than the QCD anomaly, 
{\it whatever its microscopic origin is}, can be suppressed
enough for the axion mechanism to work.

In sect. VI, 
we discuss the phenomenological viability of such a  large
radius.  It crucially depends upon 
the quantized coefficients of the K\"{a}hler moduli
superfields $T_I$ in
gauge kinetic functions which can be determined either  
by the cohomology class of vacuum configuration \cite{vafa} or by
heterotic string one-loop computation \cite{kaplu1}.
We note that 
the required large radius 
is allowed only in a rather limited class of models.
For example, for supersymmetry-preserving  compactifications
on a smooth Calabi-Yau manifold with a
vanishing $E^{\prime}_8$ field strength,
it is allowed only when
(i) the quantized flux of the
antisymmetric tensor field in $M$-theory has the  minimal
nonzero value, $[G/2\pi]=1/2$ in the notation of ref. \cite{witten3},
and (ii) the hidden gauge group $E_8^{\prime}$ is broken
by Wilson lines to a subgroup with small values of the
second Casimir $C_2={\rm tr}(T^2_{\rm adj})$.

One of the difficulties  of solving the strong CP problem by
string-theoretic axions would be a  
too large  cosmological axion mass density associated with
the axion misalignment  $\delta a\gg 10^{12}$ GeV in
the early universe \cite{pww,choi}.
Several  mechanisms have been suggested to ameliorate
this difficulty, for instance
an entropy production after the QCD phase transition
in the early universe \cite{stein,lps} or a dynamical relaxation of
the axion misalignment \cite{dvali,bank3}.
In sect. VI, we stress that 
the large compactification radius 
allows the QCD axion in $M$-theory 
to be cosmologically viable in the presence
of a late time entropy production 
without assuming any
significant suppression of the axion misalignment. 
More explicitly, we find the axion decay constant
$v\ap 10^{16}$ GeV in realistic
$M$-theory limit,
for which the QCD axion  
is cosmologically safe if there is an 
entropy production  with the
reheat temperature $T_{RH}\ap 6$ MeV which saturates the
lower bound 
from the big-bang nucleosynthesis.
We finally  discuss the  
cosmology of  the QCD axion in $M$-theory
in the case that  there is an accidental  axion-like
field whose  decay constant is much smaller than $10^{16}$ GeV.
It is pointed out  that, due to its high energy potential,
such an accidental axion is not so helpful for 
ameliorating the cosmological difficulty of the original
QCD axion in $M$-theory.

\section{Axions, couplings and scales in M-theory}

In compactified heterotic string theory,
axions
appear as the massless modes of the second rank antisymmetric
tensor field \cite{witten}:
$$
B=b_{\mu\nu}dx^{\mu}dx^{\nu}+b_I\omega^I_{i\bar{j}}dz^i
d\bar{z}^j,
$$
where $x^{\mu}$ and $z^i$ denote the coordinates
of the flat Minkowski spacetime and the internal six manifold 
respectively.
Here $b_{\mu\nu}$ is   the so-called model-independent axion, while
$b_I$ are the model-dependent axions
associated with the harmonic $(1,1)$ forms $\omega^I_{i\bar{j}}$
($I=1,...,h_{1,1}$)
on the compact (complex) six manifold.
Such axions remain  as massless modes even in strong 
string coupling
limit, i.e. $M$-theory limit, whose low energy limit can
be described by eleven-dimensional
supergravity on a manifold with boundary
\cite{horava}.
In this limit,  axions  appear as the massless modes
of the third rank antisymmetric tensor \cite{bank}:
$$
C=b_{\mu\nu}dx^{11}dx^{\mu}dx^{\nu}
+b_I\partial_{11}\omega^I_{i\bar{j}}dx^{11}dz^id\bar{z}^j.
$$

In four-dimensional effective supergravity,
the model-independent axion $b_{\mu\nu}$  
can be identified as
the pseudoscalar 
component of the dilaton superfield $S$ after the duality transformation, 
while the 
model-dependent axions $b_I$ correspond to
the pseudoscalar components
of the K\"{a}hler-moduli superfields
$T_I$. To be definite, we normalize $S$ and $T_I$ so that
their
axion components are periodic fields as:
$$
{\rm Im}(S)\equiv {\rm Im}(S)+1, \quad
{\rm Im}(T_I)\equiv {\rm Im}(T_I)+1.
$$
This would allow us to define the discrete gauge
Peccei-Quinn symmetries as
\be
Z_S: \quad S\rightarrow S+i, \quad Z_T: \quad T_I\rightarrow T_I+i.
\ee
In this normalization, 
the world-sheet sigma model action 
is given by \cite{dine}
\be
S_{WS}=\frac{1}{4\pi\alpha^{\prime}}\int d^2z \,
[4\pi^2T_I\omega^I_{i\bar{j}}\partial
X^i\bar{\partial}X^{\bar{j}}+
4\pi^2T^*_I\omega^I_{i\bar{j}}\bar{\partial}X^i\partial X^{\bar{j}}]
\ee
where the harmonic (1,1) forms $\omega^I$ span
the integer $(1,1)$ cohomology group of the target space,
viz
$$\int_{\Sigma_J}\omega^I=
i\int_{\Sigma_J} d^2z  \, \omega^I_{i\bar{j}}(\partial X^i
\bar{\partial}X^{\bar{j}}-\bar{\partial}X^i\partial X^{\bar{j}})
=2\alpha^{\prime} \delta_{IJ}.
$$ 
Note that $S_{WS}\equiv S_{WS}+2\pi$  correctly leads to
the periodicity $T_I\equiv T_I+i$.
The K\"{a}hler form of the target space is
given by $\omega=4\pi^2 {\rm Re}(T_I)\omega^I$, leading to
the internal  space volume  
$$
V_6=\frac{1}{3!}
\int \omega\wedge\omega\wedge\omega\ap \frac{1}{6}
(4\pi^2 {\rm Re}(T))^3 (2\alpha^{\prime})^3,
$$
where 
the internal six manifold is assumed  to be isotropic so that
${\rm Re}(T_I)\simeq {\rm Re}(T)$.

Using the heterotic string and $M$-theory relations
for the ten-dimensional gauge and gravitational couplings \cite{gross,horava}
\begin{eqnarray}
&&g_{10}^2=e^{2\phi}(2\alpha^{\prime})^3=2\pi (4\pi)^{2/3} l_{11}^6,
\nonumber \\
&&\kappa_{10}^2=\frac{1}{4}e^{2\phi}(2\alpha^{\prime})^4
=l^9_{11}/2R_{11},
\nonumber
\end{eqnarray}
it is now straightforward to find \cite{bank}
\begin{eqnarray}
&&\frac{1}{4\pi}{\rm Re}(S)=\frac{1}{g^2_{GUT}}=e^{-2\phi}{V_6\over
(2\alpha^{\prime})^3}
=\frac{1}{2\pi(4\pi)^{2/3}}\left(R_6\over l_{11}\right)^6,
\nonumber \\
&&4\pi^2 {\rm Re}(T)=6^{1/3}\frac{R_6^2}{2\alpha^{\prime}}
=6^{1/3}\pi (4\pi)^{2/3}
\left(R_{11}\over l_{11}\right)\left(R_6\over l_{11}\right)^2,
\end{eqnarray}
where $g_{GUT}$, 
$e^{\phi}$,
and $\kappa_{11}^2=l_{11}^{9}$ denote the four-dimensional
gauge coupling,
the heterotic string
coupling, and 
the eleven-dimensional gravitational coupling, respectively,
$R_6=V_6^{1/6}$ is the radius of the internal six manifold,
and finally $R_{11}$ is the length of the 11-th interval. ($R_{11}=\pi
\rho$ in Witten's notation \cite{witten1}.)

Obvioulsy ${\rm Re}(T)$ corresponds to
the compactification radius-squared in the heterotic string unit.
As we will argue in sect. V, 
the strong CP problem can be solved by string-theoretic axions if
the compactification radius
is large enough to yield
${\rm Re}(T) \ger 20$.
For such a large radius, 
\begin{eqnarray}
&&e^{2\phi}\ap g^2_{GUT} [2\pi^2{\rm Re}(T)]^3\ger  3\times 10^7,
\nonumber \\
&&R_{11}/R_6\ap
6^{-1/3} (2\pi)^{-1/2} g_{GUT} {\rm Re}(T)\ger 3,
\end{eqnarray}
and thus heterotic strings are so strongly coupled
that the size of the 11-th dimension is even bigger than
that of the other six dimensions.

Using the $M$-theory expression of the
four-dimensional Planck scale \cite{witten1} 
$$
M_P^2\ap 2R_{11}V_6/l_{11}^9\ap (2.4 \times 10^{18} \, {\rm GeV})^2,
$$
the mass scales in compactified $M$-theory are 
estimated in terms of 
$g_{GUT}$ and ${\rm Re}(T)$ as: 
\bea
\frac{1}{l_{11}}&\ap& 8\times 10^{16} \left(g_{GUT}\over 0.7\right)^{2/3}
\left(20\over {\rm Re}(T)\right)^{1/2} \, {\rm GeV},
\nonumber \\
\frac{2\pi}{R_6}&\ap& 2.5\times 10^{17} \left(g_{GUT}\over 0.7\right)\left(
20\over {\rm Re}(T)\right)^{1/2} \, {\rm GeV},
\nonumber \\
\frac{2\pi}{R_{11}}&\ap& 8\times 10^{16} \left(20\over{\rm Re}(T)\right)^{3/2}
\, {\rm GeV}.
\eea 
Note that for  $R_6=V_6^{1/6}$ and the length $R_{11}$ of the 11-th
interval, 
the characteristic Kaluza-Klein masses are given in the unit
of $2\pi/R_6$ and $2\pi/R_{11}$.
It was pointed out in refs. \cite{witten1,kaplu} that there is a severe
limitation on the large radius compactification even in $M$-theory
limit.
In ref.\cite{witten1}, it appeared as a lower limit on
the Newton's constant $G_N$ for a fixed value of the 
Kaluza-Klein  scale $M_{KK}\ap 2\pi/R_6$, while in ref.
\cite{kaplu} it appeared as a lower limit on 
$M_{KK}$ for a fixed value of the Planck scale
$M_P=\sqrt{1/8\pi G_N}$.
As we will discuss in sect. VI,  in our notation
this can be translated into a statement that
${\rm Re}(T)$ {\it can not} be
significantly bigger than $20$ in order for 
the four-dimensional
gauge coupling constant $g_{GUT}$ to have a realistic value.
Thus roughly speaking, what we need for the axion mechanism in $M$-theory 
is  ${\rm Re}(T)\ap 20$.
Note that  for this value of  ${\rm Re}(T)$, all  typical mass
scales in $M$-theory are much higher than the dynamical scale
of four-dimensional supersymmetry breaking which is for instance 
given by  $\Lambda_{SB}
\ap (m_{3/2}M_P^2)^{1/3}\ap 10^{13}$ GeV
in non-renormalizable hidden sector models \cite{kaplan}
with the weak scale gravitino mass $m_{3/2}\ap 10^2\sim 10^3$ GeV.

\section{Gauge kinetic functions in compactified $M$-theory}

In this section, we discuss some model-independent features
of holomorphic gauge kinetic functions in the effective four-dimensional
supergravity
models which correspond to the low energy limit of
compactified $M$-theory. 
The discrete gauge symmetries $Z_{S,T}$ of Eq. (2) 
imply that in the limit of ${\rm Re}(S)\gg 1$ and ${\rm Re}(T_I)\gg 1$,  
gauge kinetic functions can be written as 
\be
4\pi f_a= k_a S + \frac{1}{2}l_{a I} T_I + \triangle_a,
\ee
where $k_a$ and $l_{aI}$ are quantized real coefficients, and 
$\triangle_a$
denote the piece of order unity (or less)
which is independent of $S$ and $T_I$
or the piece which is  suppressed by  $e^{-2\pi S}$ or $e^{-2\pi T_I}$.
Here  gauge kinetic functions are normalized
as ${\rm Re}(f_a)=1/g_a^2$ and ${\rm Im}(f_a)
=\theta_a/8\pi^2$ where $g_a$ and $\theta_a$
denote the four-dimensional gauge couplings and the Yang-Mills 
vacuum angles,
respectively. 

The quantized coefficients  $k_a$ and $l_{aI}$ are  
{\it unchanged} when one moves from the $M$-theory domain
to the  domain of weakly coupled heterotic string in the
moduli space of the theory, and thus
they can be determined within the 
weakly coupled heterotic string theory.
Note that in the region of  ${\rm Re}(S)\gg 1$
and ${\rm Re}(T_I)\gg 1$ where the expression (7) is
valid,  the heterotic  string
can be weakly coupled, 
 $e^{2\phi}\ap 32 \pi^7{\rm Re}(T)^3/{\rm Re}(S)\ll 1$, 
in the domain where
the four-dimensional
gauge couplings are small enough, $g^2_{GUT}\ap 4\pi/{\rm Re}(S)
\ll 1/ (2\pi^2 {\rm Re}(T))^3$.
(Of course, $e^{2\phi}\gg 1$
in the domain giving a realistic value of $g^2_{GUT}\ap \frac{1}{2}$.)
In the weak string coupling limit where  heterotic
string perturbation theory is a good approximation,
$k_a$ can be identified (for non-Abelian gauge groups)
as  the level of
the Kac-Moody algebra whose zero modes generate the $a$-th
gauge boson, and thus are positive integers. 

If the compactification is simple enough, e.g. orbifolds,
one may determine $l_{aI}$  by computing the 
string one-loop threshold  correction to  gauge kinetic functions 
\cite{kaplu1}.
In fact, one can extract  some model-independent 
information on $l_{aI}$
without resorting to 
any string loop calculation.
Under the discrete gauge transformation $Z_T: T_I\rightarrow T_I+i$,
the Yang-Mills vacuum angles
transform as 
$$
\theta_a\rightarrow \theta_a+\pi l_{aI},
$$
and thus $\theta_a$ and $\theta_a+\pi l_{aI}$ are required
to be physically equivalent.
Applying to the usual
$2\pi$ periodicity relation $\theta_a\equiv \theta_a + 2\pi$,
this may be considered to imply that $l_{aI}/2$ are integers,
however this is not necessarily the case in string theory.
Due   to the existence of the model-independent axion,
in string theory
the vacuum angles enjoy an additional equivalence
relation:
\be
\theta_a\equiv \theta_a+2\pi k_a\gamma,
\ee
where $\gamma$ is an arbitrary real constant.
We are then led to
\be
\frac{1}{2}l_{aI}-k_a\gamma={\rm integer},
\ee
and as a consequence 
\be
\frac{1}{2}(k_b l_{aI}-k_a l_{bI})={\rm integer}.
\ee

The equivalence relation (8)
can be most easily understood by treating the model-independent
axion as an antisymmetric tensor field $b=b_{\mu\nu}dx^{\mu}\wedge dx^{\nu}$ 
whose gauge invariant field strength 
is given by $H=db-\sum_a k_a\omega^a_{YM}+\omega_L$
where $\omega^a_{YM}$ and $\omega_L$ denote the Yang-Mills
and Lorentz Chern-Simon
three forms, respectively \cite{witten2}.  For the gauge invariance of $H$,
only the large
gauge transformations  under which $\int_{R^3}\sum_a k_a\omega^a_{YM}
=\sum_a k_a n_a$
is invariant are allowed in the theory. 
For such large gauge transformation $U$, we have
$\sum_a k_a(n_a-n^{\prime}_a)=0$ where
$|n^{\prime}_a\rangle=U|n_a\rangle$.
Then for the $\theta$-vacuum state defined as 
$|\theta_a\rangle=\sum_{n_a}\exp(i\sum_a n_a\theta_a)|n_a\rangle$, we
have the equivalence relation
$$
U|\theta_a\rangle
=U|\theta_a+k_a\gamma\rangle,
$$
and thus the equivalence relation (8) also.
In the dual pseudo-scalar description
of the model-independent axion which we are using here,
the  equivalence relation (8) 
is nothing but to make one combination
of the vacuum angles to be gauged away by the shift of the model-independent
axion  
${\rm Im}(S)$. 
A property which distinguishes ${\rm Im}(S)$
from other axion-like fields is that
its non-derivative couplings
appear {\it always} through the combination ${\rm Im}(f_a)\propto
k_a{\rm Im}(S)+\frac{1}{2}l_{aI}{\rm Im}(T_I)$ whose vacuum value
corresponds to the vacuum angle $\theta_a$.
(This amounts to the dual property  of the gauge-invariance
of $H=db+\omega_L-\sum_ak_a\omega^a_{YM}$ in the antisymmetric
tensor formulation.)
Due to this special property of ${\rm Im}(S)$,
vacuum angles  related by a shift of ${\rm Im}(S)$
are physically equivalent to each other. 
Model-dependent
axions ${\rm Im}(T_I)$  have other type of
nonderivative couplings in addition to the couplings
through ${\rm Im}(f_a)$, e.g. those 
induced by world-sheet instantons and possibly others,
and thus do not provide additional equivalence relation
for the vacuum angles. 
However they can still be useful
for the strong CP problem if their non-derivative couplings
other than the QCD anomaly are suppressed enough.

Let us now  focus on 
the $E_8\times E^{\prime}_8$ 
theory compactified on a large smooth Calabi-Yau manifold and let
$f_{E_8}$ and $f_{E^{\prime}_8}$ denote
the visible ($\subset E_8$) and  hidden ($\subset E^{\prime}_8$)
sector gauge kinetic functions, respectively.
In this case, $k_a=1$ and thus 
\bea
4\pi f_{E_8} &=& S+\frac{l_I}{2}T_I+\triangle_{E_8}, \nonumber \\
4\pi f_{E_8^{\prime}} &=& S+\frac{l^{\prime}_I}{2}T_I+
\triangle_{E^{\prime}_8}.
\eea
The coefficients  $l_{I}$ and $l^{\prime}_I$ can  be 
determined in the limit
of ${\rm Re}(S)\gg 1$ and ${\rm Re}(T)\gg 1$ while $e^{2\phi}\ll 1$,
in which
ten-dimensional effective field theory provides a good approximation.
In this limit, the anomaly cancellation mechanism 
implies that the axion couplings of ${\rm Im}(T_I)$
are  entirely due to 
the Green-Schwarz term in 
ten-dimensional  field theory \cite{choi1,ibanez,bank}:
$$
S_{GS}=\frac{1}{288 
 (2\pi)^5}\int B [{\rm Tr}(F^4
)-\frac{1}{300}
({\rm Tr}(F^2))^2
-\frac{1}{10}{\rm Tr}(F^2)
{\rm tr}(R^2)+...].
$$
For the compactification on $M_4\times M_6$,
the above Green-Schwarz term leads to the following
axion couplings in the four-dimensional effective
theory:
\be
\frac{1}{32\pi}
\int_{M_6} \omega^I\wedge I_4
\int_{M_4}{\rm Im}(T_I)[{\rm tr} (F^{\mu\nu}\tilde{F}_{\mu\nu}
-F^{\prime\mu\nu}\tilde{F}^{\prime}_{\mu\nu})],
\ee
where $F_{\mu\nu}$ and $F^{\prime}_{\mu\nu}$ denote the
$E_8$ and $E^{\prime}_8$ field strengths on $M_4$,  $\tilde{F}_{\mu\nu}$
and $\tilde{F}^{\prime}_{\mu\nu}$ are their duals, and
the four-form $I_4$ is defined as 
$$
8\pi^2 I_4={\rm tr} (F\wedge F)-\frac{1}{2}{\rm tr}(R\wedge R), 
$$
with $F=\frac{1}{2}F_{AB}dy^A\wedge dy^B$ and $R=\frac{1}{2}R_{AB}
dy^A\wedge dy^B$, where $y^A$ correspond to the real coordinates
of $M_6$.
Here we set $2\alpha^{\prime}=1$,
${\rm tr} F^2$ is 1/30 of the trace over the
$E_8$ adjoint representation, and 
$B={\rm Im}(\tilde{\omega})=4\pi^2{\rm Im}(T_I)\omega^I$
where $\tilde{\omega}$ denotes the complexified K\"{a}hler form.
To arrive at the above result, we have used  
the model-independent
constraint
$$
\int_{M_6} \omega^I \wedge (I_4+I_4^{\prime})=
\frac{1}{8\pi^2}\int_{M_6} \omega^I\wedge dH= 0,
$$
where $8\pi^2 I_4^{\prime}=[{\rm tr}(F^{\prime}\wedge
F^{\prime})-\frac{1}{2}{\rm tr}(R\wedge R)]$.
Matching the axion coupling (12) to the gauge kinetic
functions in (11), we finally  obtain
\be
l^{\prime}_I=-l_I=
\int_{M_6} \omega^I \wedge I_4.
\ee
Note that $\omega^I$ are normalized
as $\int_{\Sigma_I}\omega_J=\delta_{IJ}$ (in the unit
of $2\alpha^{\prime}=1$),
and then  $\frac{1}{2}(l_I-l^{\prime}_I)=l_I$ are 
integral as required by (10).
The relation $l_I=-l^{\prime}_I$ was 
noted before by several authors \cite{choi1,ibanez,bank} 
for the case of (2,2)  Calabi-Yau compactifications.
Here we stress that  it is 
valid for generic smooth compactifications and thus is a
rather
model-independent prediction of string theory.
(It has been noted recently that 
$l_I=-l^{\prime}_I$ also in some orbifold models
\cite{nilles}.)

As was noted by Banks and Dine \cite{bank},
the weak-coupling result (13) can be confirmed by the Witten's
strong-coupling expansion in $M$-theory
\cite{witten1}, which leads to 
$$
V_6(E_8^{\prime})-V_6(E_8)=2\pi (4\pi)^{-2/3} R_{11} l_{11}^3
\int_{M_6} \omega\wedge I_4,
$$
where  $V_6(E_8)$ and $V_6(E^{\prime}_8)$
denote the Calabi-Yau volume on the $E_8$-boundary 
and the $E^{\prime}_8$-boundary,  respectively,
and $\omega=4\pi^2{\rm Re}(T_I)\omega^I$ is the K\"{a}hler form.
With the relation (4), this amounts to
\be
4\pi {\rm Re}(f^{\prime}_{E_8})-4\pi {\rm Re}(f_{E_8})
={\rm Re}(T_I) \int_{M_6} \omega^I\wedge I_4,
\ee
which  gives   
$\frac{1}{2}(l^{\prime}_I-l_I)=\int_{M_6} \omega^I\wedge I_4$.
The above result  
indicates also  that $\triangle_a$
in the gauge kinetic function
corresponds to  higher order correction 
in the strong coupling expansion.
Note that, in the $M$-theory limit under consideration,
 ${\rm Re}(T_I)\gg 1$, while
${\rm Re}(\triangle_a)$ are essentially
of order unity or less.
Also in $M$-theory,
$I_4$ and $I^{\prime}_4$ can be identified
as the boundary values of the field strength $G=dC+...$ of the three form
field $C$ in eleven-dimensional supergravity:
$\frac{1}{\pi}[G]_{E_8}=I_4$ and $\frac{1}{\pi}[G]_{E^{\prime}_{8}}=
I^{\prime}_4$ \cite{horava}, and then 
the Bianchi identity    
$\frac{1}{\pi} \int dG=
\int_{E_8} I_4+\int_{E^{\prime}_{8}} I_4^{\prime}=0$
implies  $l_I=-l^{\prime}_I$.
We then recover the
weak-coupling result (13) through the strong coupling calculation.

For supersymmetry-preserving  compactifications on a smooth
Calabi-Yau manifold,
Eq. (14) has an alternative expression.
In both the weakly coupled heterotic string theory and the
$M$-theory limits, 
the starting point 
of such compactifications 
is a  manifold of $SU(3)$ holonomy, and a holomorphic
Yang-Mills connection satisfying the K\"{a}hler-Yang-Mills equations:
$F_{ij}=g^{i\bar{j}}F_{i\bar{j}}=0$.
We then have
$\int \omega\wedge{\rm tr}(R\wedge R)=-\frac{1}{2}
\int {\rm tr}(R_{AB}R^{AB})$ and
similarly $\int \omega\wedge{\rm tr}(F\wedge F)=
-\frac{1}{2} \int {\rm tr}(F_{AB}F^{AB})$.
Also the string theory Bianchi identity
$dH=8\pi^2(I_4+I^{\prime}_4)$ (or the dilaton equation of motion) 
or its $M$-theory counterpart 
leads to the condition $\int{\rm tr}(F_{AB}F^{AB}+
F^{\prime}_{AB}F^{\prime AB }-R_{AB}R^{AB})=0$ \cite{book}.
Putting these together, we  find
\be
4\pi {\rm Re}(f_{E_8})
-4\pi {\rm Re}(f_{E^{\prime}_8})
=\sum l_I{\rm Re}(T_I)
=\frac{1}{128\pi^4}[\int_{M_6}F_{AB}F^{AB}
-\int_{M^{\prime}_6}F^{\prime}_{AB}F^{\prime AB}].
\ee
With this expression, it is easy to realize that
if the $E^{\prime}_8$ field strength does
vanish, i.e. $F^{\prime}_{AB}=0$, then $\sum l_I{\rm Re}(T_I)>0$
independently of the relative ratios ${\rm Re}(T_I)/{\rm Re}(T_J)$,
and thus at least one of $l_I$'s is a positive integer, while other $l_I$'s
are still {\it non-negative} integers.

\section{High Energy Potential of the Model-Independent Axion}

In this and the next  sections, we analyze in detail  the axion potential
due to  the explicit breakings other than the QCD anomaly
of the non-linear $U(1)_{PQ}$
symmetries of string-theoretic axions. 
As is well known, 
in order for the strong CP problem to be solved by the axion
mechanism, one needs an anomalous
global $U(1)_{PQ}$ symmetry whose breaking
other than the QCD anomaly is so tiny that its contribution 
to the axion potential satisfies
\be
\delta V_{\rm axion}\ler 10^{-9} f_{\pi}^2m_{\pi}^2\ap 10^{-13} \,
{\rm GeV}^4.
\ee
To examine the possibility of such a $U(1)_{PQ}$ symmetry, 
we start from the non-linear $U(1)_S$ symmetry 
associated with the model-independent
axion ${\rm Im}(S)$,
$$
U(1)_S: \quad S \rightarrow S+i\alpha,
$$
where $\alpha$ is a continuous real parameter.
Since $k_a\neq 0$ in both the observable sector and
the hidden sector gauge kinetic functions written as (7),
$U(1)_S$ is broken not only by the QCD anomaly, but also by
the hidden sector gauge anomaly.
Our primary concern  in this section is to  examine 
the possibility that  the 
high energy  potential $\delta V_{{\rm Im}(S)}$
of the model-independent axion due to 
the hidden sector anomaly is suppressed enough to
satisfy the bound (16).

The axion potential  
$\delta V_{{\rm Im}(S)}$ 
is somewhat sensitive to how four-dimensional supersymmetry is spontaneously 
broken. 
As we will see in sect. VI,
${\rm Re}(T)$ can {\it not} be significantly bigger than $20$, 
and then the typical  mass scales in $M$-theory
estimated in (6) are  too large to be identified as the dynamical
scale of four-dimensional supersymmetry breaking.
It is then the most attractive possibility that
the supersymmetry breaking scale 
is small compared to the typical $M$-theory  scale, 
say $2\pi R_6^{-1}$ or $2\pi R_{11}^{-1}$ which is about  $10^{17}$ GeV,
since it is induced by
nonperturbative effects like the hidden sector gaugino (matter)
condensations (or some $M$-theoretical nonperturbative
effects).
In this scheme, after integrating out the hidden sector gauge and
matter multiplets,
the resulting effective superpotential 
can be written as
\be
W_{\rm eff}=W_1+W_2+W_{\rm obs}=\Omega_1 e^{-2\pi\gamma_1 S}+\Omega_2 
e^{-2\pi\gamma_2 S}+W_{\rm obs},
\ee
where $\Omega_1$ and $\Omega_2$ depend upon neither
$S$ nor the observable sector fields (but generically they depend
upon the other moduli), and
$\gamma_1$ and $\gamma_2$ are {\it different} rational coefficients.
Here $W_1=\Omega_1 e^{-2\pi\gamma_1 S}\ap m_{3/2}M_P^2$ 
is the leading term in $W_{\rm eff}$,
while $W_2=\Omega_2 e^{-2\pi\gamma_2 S}$ is the next-to-leading term
when the observable sector
fields are set to zero, and finally $W_{\rm obs}$
includes the terms depending
upon the observable sector fields.

Let us first consider the hidden sector contribution to 
$\delta V_{{\rm Im}(S)}$.
From the supergravity potential 
\be
V_{\rm SG}= e^{K_{\rm eff}/M_P^2}[K_{\rm eff}^{IJ}
D_IW_{\rm eff}(D_JW_{\rm eff})^*-3|W_{\rm eff}|^2/M_P^2]+
({\rm D \,\, terms}),
\ee
it is easy to find
\be
\delta V_{{\rm Im}(S)}\ap W_1W_2^*/M_P^2
\ap (W_2^*/W_1) m_{3/2}^2 M_P^2,
\ee
where $K_{\rm eff}$ is the effective K\"{a}hler potential after integrating
out the hidden sector gauge and matter multiplets.
Obviously, in order for this axion potential 
to satisfy the bound (16), one needs a huge
hierarchy between the leading term $W_1$ and the next-to-leading
term $W_2$: 
$$
\frac{W_2}{W_1}\ap \frac{W_2}{m_{3/2}M_P^2} \ler
10^{-55}\left({\rm TeV}\over m_{3/2}\right)^2.
$$

If there are more than one non-Abelian hidden sector gauge
groups which would yield 
multi-gaugino (matter) condensations,
$W_2$ receives a contribution from the second largest condensate.
This contribution is typically
much bigger than $\Lambda_{QCD}^3$.
In particular, if the dilaton is stabilized 
by the racetrack mechanism  \cite{race}, 
$W_2$ and $W_1$ are comparable 
to each other, yielding $\delta V_{{\rm Im}(S)}\ap m_{3/2}^2 M_P^2$. 
We thus conclude that in models with multi-gaugino (matter)
condensations,
the model-independent axion receives a harmful
high energy potential much bigger than $10^{-9}f_{\pi}^2m_{\pi}^2$.

If there is only  a single non-Abelian hidden sector gauge group,
higher dimensional operators would
be  responsible
for the next-to-leading  term $W_2$ in the effective
superpotential (17).
(One may suffer from the dilaton runaway in this case.
Here we assume as in ref. \cite{bank1} that  the dilaton
is stabilized 
by a large $U(1)_S$-preserving nonperturbative correction
to the K\"{a}hler potential.)
As an example, let us consider
a typical hidden sector with $SU(N_c)$ gauge group,
$N_f$ quark flavors $(Q+Q^c)$ and $N_s$ singlets $A$, and
the tree level superpotential 
$$
W_{\rm tree}=A^3+AQQ^c+...,
$$ 
where the Yukawa couplings of order
unity are omitted and the ellipsis denotes higher dimensional
operators.  We also assume $k_a=1$ for the hidden sector
gauge kinetic function written as (7).
We then have  
$$
W_1\ap \langle W^a_{\rm hid}W^a_{\rm hid}\rangle
\ap \Lambda_{\rm hid}^3, \quad
\langle A\rangle\ap \Lambda_{\rm hid}, \quad
\langle QQ^c\rangle\ap 
\Lambda_{\rm hid}^2,
$$
where  $W^a_{\rm hid}$ is the chiral superfield whose lowest
component corresponds to  the hidden sector gaugino,
and  $\Lambda_{\rm hid}\propto e^{-2\pi S/(3N_c-N_f)}$ is the 
dynamical scale of the hidden sector gauge interaction.
In this case, the next-to-leading
term  $W_2$ can be induced by
gauge-invariant non-renormalizable operators of the form 
\be
M_P^3\int d^2\theta \left(W^a_{\rm hid}W^a_{\rm hid}\over M_P^3\right)^n
\prod_{k=1}^m\left(\Phi_{I_k}\over M_P\right),
\ee
where $\Phi_I$ denote the hidden matter superfields: $\Phi_I=
(A, Q, Q^c)$.  
Obviously the above operator gives rise to 
$$
W_2\ap \frac{\Lambda_{\rm hid}^{3n+m}}{M_P^{3n+m-3}}\ap
M_P^3\left(W_1\over M_P^3\right)^{(3n+m)/3}.
$$
and thus
\be
\delta V_{{\rm Im}(S)}\ap \frac{W_1W_2^*}{M_P^2}\ap
m_{3/2}^2M_P^2\left(m_{3/2}\over M_P\right)^{(3n+m-3)/3}.
\ee
In order for this axion potential  to satisfy the bound (16), 
we need 
\be
3(n-1)+m\ger \frac{380+6\ln(m_{3/2}/{\rm TeV})}{35-\ln (m_{3/2}/{\rm TeV})},
\ee 
i.e. all   non-renormalizable operators 
whose mass dimension $d=3n+m+1\ler 14$  have to be forbidden.

It has been pointed out in  \cite{bank1} that, if there is no hidden matter,
a simple discrete gauge symmetry can eliminate
all dangerous non-renormalizable operators.
The  example considered in \cite{bank1} is a discrete gauge
$R$ symmetry $Z_5$ under which 
$d^2\theta\rightarrow
e^{-i2\pi/5} d^2\theta$.
The model-independent axion is  transformed also  as
$S\rightarrow S+ iN_c/5$
to cancel the $Z_5\times SU(N_c)\times SU(N_c)$ anomaly.
In the absence of any hidden matter multiplet,
this $Z_5$ allows only the operators  with $n\geq 6$ in (20),
thereby satisfying the condition (22).
However, if the hidden sector contains
matter multiplets, it becomes much more nontrivial to fulfill
the condition (22). The $Z_5$-charges of
the matter multiplets have to be judiciously chosen 
in order to forbid all dangerous
non-renormalizable operators with the mass dimension $d\ler 14$.

Even when the hidden sector is  adjusted to make its contribution
to $\delta V_{{\rm Im}(S)}$  satisfy the bound (16),
one still has to carefully tune the observable sector
to avoid a harmful axion potential from the observable sector
dynamics.
Let us suppose that the hidden sector is tuned to have $W_2/W_1\ler
10^{-55}$ by means of a discrete gauge symmetry, e.g.  $Z_5$
of ref. \cite{bank1}.
To see how a sizable
$\delta V_{{\rm Im}(S)}$  can still be induced,
let us consider a model whose  observable sector 
contains a matter
multiplet $\phi$ which is neutral  under unbroken
continuous gauge symmetries, but carries  a nonzero
$Z_5$-charge $q_{\phi}$: 
$$
Z_5: \quad 
d^2\theta \rightarrow e^{-i2\pi/5} d^2\theta, \quad
\phi\rightarrow e^{i2\pi q_{\phi}/5}\phi.
$$
Note that such a matter multiplet appears quite often in
compactified string models.
Although $\phi$ does not have any renormalizable
coupling with the hidden sector fields, it can still couple
to the hidden sector via non-renormalizable operators.
For instance, gauge symmetries including
$Z_5$ allow the couplings
of the form
\be
M_P^3\int d^2\theta \,  \left[\left(
\frac{W^a_{\rm hid}W^a_{\rm hid}}{M_P^3}\right)^{N}
\left(\frac{\phi}{M_P}\right)^{M}
+\left(\frac{\phi}{M_P}\right)^{L}\right],
\ee
and thus
the folllowing effective superpotential
$$
W_{\rm eff}=\Omega_1 e^{-2\pi\gamma_1 S}+
M_P^{3(1-N)}\Omega_1^{N}e^{-2\pi N
\gamma_1 S}\left(\phi\over M_P\right)^{M}
+\left(\phi\over M_P\right)^{L}.
$$
for appropriate values of the positive integers
$N$, $M$, and $L$.
To be more specific, let us consider the case (I) with 
$q_{\phi}=-1$ for which  $N=1$, $M=5$,
$L=4$, and also the case (II) with $q_{\phi}=2$
for which  $N=2$, $M=2$, $L=3$.
Then the supergravity potential (18) contains 
\[
\delta V_{SG}\ap 
\left\{ \begin{array}{ll}
(m_{3/2}^*\phi^{4}/M_P)+(|m_{3/2}|^2\phi^{5}/M_P^3) & 
\quad \mbox{: Case (I)} \\
(m_{3/2}^*\phi^{3})+(|m_{3/2}|^2m_{3/2}\phi^2/M_P)
& \quad \mbox{: Case (II)}
\end{array}
\right. \] 
where the model-independent axion 
${\rm Im}(S)$ appears through the complex gravitino mass
defined as  
$m_{3/2}\ap W_1/M_P^2=\Omega_1 e^{-2\pi\gamma_1 S}/M_P^2$.
Although it appears to depend upon ${\rm Im}(S)$,
the first term of the above supergravity potential in
each case does {\it not}
contribute to the true axion potential since their
${\rm Im}(S)$-dependence can be eliminated by
the field redefinitions:  
$\phi\rightarrow e^{-i\pi \gamma_1 {\rm Im}(S)/2}\phi$
for the case (I) and $\phi\rightarrow
e^{-i2\pi \gamma_1 {\rm Im}(S)/3}\phi$ for the case (II).
As a result, the high energy potential of the model-independent axion
in each case
is estimated to be
\bea
&& {\rm Case \, (I)}: \quad \delta V_{{\rm Im}(S)}\ap
|m_{3/2}|^2\langle \phi\rangle^5/M_P^3, 
\nonumber \\
&& {\rm Case \, (II)}: \quad \delta V_{{\rm Im}(S)}\ap
|m_{3/2}|^2m_{3/2}\langle \phi\rangle^2/M_P. 
\eea

Of course it crucially depends upon the size of  $\langle\phi\rangle$
whether the above axion potential satisfies the bound (16).
In fact, a nonzero $\langle \phi\rangle$ can lead to interesting
phenomenological consequences.
For instance, it may generate 
the $\mu$-term of the Higgs doublets $H_u$ and $H_d$
through the coupling $\int d^2 \, \theta \phi H_uH_d$ with the weak scale
vacuum value
$\langle \phi\rangle\ap m_{3/2}$, or through
the coupling $\frac{1}{M_P}\int d^2\theta \, \phi^2 H_uH_d$
with the intermediate scale vacuum value $\langle \phi\rangle=
\sqrt{m_{3/2}M_P}$.
It may also generate 
an intermediate scale mass of the right-handed neutrino $N$  through
the coupling $\int d^2\theta \, \phi NN$, leading to the small neutrino mass
$m_{\nu}\ap m_{3/2}^2/\sqrt{m_{3/2}M_P}$ via the seesaw mechanism.
In case (I), $\phi$ has a flat potential
in the limit of $m_{3/2}\rightarrow 0$ and
$M_P\rightarrow\infty$.
Including the non-renormalizable but supersymmetric potential
term $|\phi|^6/M_P^2$ 
together with the     soft mass term $m_{3/2}^2|\phi|^2$
which receives a negative radiative correction due to 
the Yukawa coupling $\int d^2\theta \, \phi NN$,
one easily  finds $\langle \phi\rangle\ap \sqrt{m_{3/2}M_P}$
for the case (I) \cite{yana}.
Also for the case (II) with the Yukawa coupling
$\int d^2\theta \, \phi H_u H_d$, a similar analysis 
yields $\langle \phi\rangle\ap m_{3/2}$.
Then in both cases (I) and (II),
the high energy potential of
the  model-independent axion $\delta V_{{\rm Im}(S)}\gg
10^{-9} f_{\pi}^2 m_{\pi}^2$
for the weak scale gravitino mass $m_{3/2}\ap 10^2\sim 10^3$ GeV.

In fact,  
$\delta V_{{\rm Im}(S)}$ is naively expected
to be of order $m_{3/2}^2M_P^2$ since it
is essentially due to hidden sector dynamics triggering
supersymmetry breaking.
As we have noted,  it is really of order $m_{3/2}^2M_P^2$
in generic cases with multi-gaugino (matter) condensations whose sizes
are comparable to each other.
In the cases that $\delta V_{{\rm Im}(S)}\ll 
m_{3/2}^2M_P^2$,
there is a simple explanation for a small $\delta V_{{\rm Im}(S)}$.
In such cases, we always have an approximate {\it accidental}
global $U(1)_X$ symmetry
whose current divergence is given by
$\partial_{\mu}J^{\mu}_X=(F\tilde{F})_1+\Gamma_X$
where $(F\tilde{F})_1$ corresponds to  the $U(1)_X\times
G_1\times G_1$ anomaly for the non-Abelian gauge group $G_1$
which is 
responsible for the leading term $W_1$ in the effective superpotential (17),
and $\Gamma_X$ stands for the other symmetry breaking terms
which are presumed to be weaker than the strongest
anomaly $(F\tilde{F})_1$.
For the models  that we have discussed above, 
we have
$$
U(1)_X: \quad \Phi\rightarrow e^{i\beta X(\Phi)}\Phi,
$$ 
where the $U(1)_X$ charges 
are given by $X(d^2\theta)=-1$,
$X(A)=1/3$, $X(QQ^c)=2/3$, and
$X(\phi)=1/4$ ($1/3$) for the case (I) (the case (II)),
and then $\Gamma_X$ includes the $U(1)_X$-breakings
by the non-renormalizable operators in (20) and (23).
For $k_a=1$, the model-independent axion has a universal coupling
to the gauge anomalies and then the  $U(1)_S$ current $J^{\mu}_S
\propto \partial^{\mu}{\rm Im}(S)$ obeys
$\partial_{\mu}J^{\mu}_S=\sum_a(F\tilde F)_a+(R\tilde{R})$
where $(R\tilde{R})$ denotes the irrelevant gravitational chiral anomaly.
In the presence of $U(1)_X$,
the model-independent axion can be identified as 
the pseudo-Goldstone boson of 
\be
U(1)_{S-X}: \quad S\rightarrow S+ik\beta, \quad \Phi\rightarrow 
e^{i\beta X(\Phi)}\Phi, 
\ee 
where the coefficient $k$ is chosen to make $U(1)_{S-X}$
to be free from the strongest hidden
sector anomaly $(F\tilde{F})_1$.
If the hidden sector gauge group is semisimple,
$U(1)_{S-X}$ is still broken by the second strongest hidden sector
anomaly, leading to 
$\delta V_{{\rm Im}(S)}\ap W_1W_2^*/M_P^2\gg 10^{-9}f_{\pi}^2m_{\pi}^2$
where $W_2$ is induced by the second strongest hidden sector
gauge interaction.
In the case that $G_1$ is the only non-Abelian hidden sector gauge
group, major explicit breaking of $U(1)_{S-X}$ is due to
the non-renormalizable
operators (20) and (23).
The axion potential $\delta V_{{\rm Im}(S)}$ is then suppressed
by  an insertion
of these small $U(1)_{S-X}$-breaking operators as can be seen in (21) and (24).
Although the suppression was not enough
so that  $\delta V_{{\rm Im}(S)}\gg 10^{-9}f_{\pi}^2m_{\pi}^2$
in our examples,
there may exist a compactified
string model with an accidental $U(1)_X$ which is 
good  enough to yield $\delta V_{{\rm Im}(S)}\ler 10^{-9} f_{\pi}^2
m_{\pi}^2$.
However this possibility is too much model-dependent
and implementing this scenario in the context of  string/$M$-theory
appears to be quite nontrivial.

In summary, in this section
we have examined the possibility that the
high energy potential $\delta V_{{\rm Im}(S)}$
of the model-independent axion
due to the hidden sector anomaly
is suppressed enough to be smaller than $10^{-9}f_{\pi}^2m_{\pi}^2$,
thereby solving the strong CP problem by the model-independent
axion alone.
First of all, 
this appears to be {\it not} possible 
if the hidden sector dynamics 
yields multi-gaugino (matter)
condensates.
Even in the case that there is only one non-Abelian
hidden sector gauge group,
it requires a careful tuning of
both the hidden sector and the observable sector
to forbid all dangerous higher dimensional
operators. This is equivalent to having
an accidental global $U(1)_X$ for which the 
combination $U(1)_{S-X}$ of Eq. (25) which is
designed to be free from the hidden
sector anomaly  
is  so good a  symmetry that the corresponding high energy
axion potential 
is smaller than $10^{-9}f_{\pi}^2m_{\pi}^2$.
Arranging the observable sector to have such $U(1)_X$
appears to be highly nontrivial.
At any rate,
our study in this section shows that
it is much more nontrivial 
than what has been suggested 
in the previous works \cite{bank1,wu} to 
solve the strong CP problem by
the model-independent axion alone.

\section{Peccei-Quinn symmetry in
the large radius limit}

In $M$-theory limit where the world-sheet (membrane) instanton
effects are highly suppressed, the desired $U(1)_{PQ}$ symmetry satisfying
the bound (1) may appear as a linear combination
of the nonlinear $U(1)$ symmetries of the model-independent
axion and the model-dependent K\"{a}hler axions \cite{bank}.
To examine this possibility,
let us consider a model compactified
on a smooth Calabi-Yau manifold with 
$h_{1,1}=1$,
and also assume that there is only one non-Abelian
hidden  sector gauge group from $E_8^{\prime}$.
It is rather straightforward to generalize our discussion
to more general cases that $h_{1,1}>1$ and/or the hidden sector
gauge group is semi-simple.
Our starting point is the  visible and hidden sector  gauge kinetic
functions $f_{E_8}$ and $f_{E^{\prime}_8}$
in the limit 
of ${\rm Re}(S)\gg 1$ and 
${\rm Re}(T)\gg  1$:
\begin{eqnarray}
4\pi f_{E_8}&=&S+\frac{l}{2} T+\triangle_{E_8},
 \nonumber \\
4\pi f_{E_8^{\prime}}&=&S-\frac{l}{2} T +\triangle_{E_8^{\prime}},
\end{eqnarray}
where $l$ is integral, and again $\triangle_a$ 
corresponds to the piece of order one 
which is 
independent of $S$ and $T$ or the piece which is suppressed by 
$e^{-2\pi S}$ or $e^{-2\pi T}$.
The above form of gauge kinetic functions naturally leads to
the following 
$U(1)_{PQ}$ symmetry
\be
U(1)_{PQ}: \quad S\rightarrow S+i\alpha, \quad 
T\rightarrow T+2i\alpha/l,                                                 ,
\ee
which is free from the hidden sector anomaly.
This $U(1)_{PQ}$ would solve the strong CP
problem if its explicit breaking other than  the QCD anomaly
is so tiny that the associated axion potential satisfies
the bound (1).
Note that $l$ is required to be nonzero in order to avoid
the hidden sector anomaly, while keeping the breaking
by the QCD anomaly.

It has been argued that generic quantum gravity effects
may break $U(1)_{PQ}$ explicitly \cite{holman}.
Although it is somewhat clear that world-sheet (membrane) instanton
effects are suppressed by $e^{-2\pi T}$ in the large radius
limit, in the absence of the full understanding
of the $M$-theory dynamics,
one may still wonder that some unknown $M$-theoretical 
effects other than world-sheet (membrane) instantons  
breaks $U(1)_{PQ}$ significantly 
even in the large radius limit.
In this regard, it would be desirable if 
$U(1)_{PQ}$ in the large radius limit 
can be protected by some gauge
symmetries at the compactification scale.
In the following, we 
argue that supersymmetry and the discrete gauge
symmetries highly constrain the possible explicit breaking
of $U(1)_{PQ}$, and as a result the potentially harmful breaking
of $U(1)_{PQ}$, whatever its microscopic  origin is, is suppressed
enough if the compactification
radius is large enough.

To proceed, let us assume that 
the discrete gauge symmetries  $Z_{S,T}$ of Eq. (2) 
are {\it not} spontaneously
broken by the $M$-theory dynamics at scales above the compactification
scale, and consider the limit
$$
{\rm Re}(T)\gg 1, \quad {\rm Re}(S)-\frac{l}{2}{\rm Re}(T)\gg 1,
$$
in which the four-dimensional gauge couplings $g_a^2={\rm Re}
(f_a)^{-1}$
at the compactification scale are small enough.
In this limit,  $U(1)_{PQ}$-breaking  
in a holomorphic operator $F$ which is
generated by the $Z_{S,T}$-preserving $M$-theory dynamics
is estimated as:
\be
\delta_{PQ}F\ap M_P^{d}  \exp [-2\pi \{p T+ q(S-lT/2)\}],  
\ee
where 
$p$ is a positive integer, while $q$ is a non-negative
integer.
Here $d$ denotes the mass dimension of $F$
and we have used the fact that all non-derivative couplings
of ${\rm Im}(S)$ are required to appear through the combinations
${\rm Im}(f_a)\propto {\rm Im}(S)\pm \frac{1}{2}l{\rm Im}(T)$.
Note that in order to be a $U(1)_{PQ}$-breaking piece, $p$ 
is required to be non-zero, while the coefficient $q$
of the $U(1)_{PQ}$-invariant combination
$(S-\frac{1}{2}lT)$ can be zero.

Again the integers $p$ and $q$ in (28) are unchanged
when one moves from the $M$-theory domain to the
domain of weakly coupled heterotic string, and thus
they can be determined within 
the weakly coupled  heterotic string theory.
For holomorphic gauge kinetic functions,  $U(1)_{PQ}$ is generically broken
by world-sheet instanton effects 
without any suppression by $e^{-2\pi S}$ 
at string one-loop order \cite{kaplu1}, and thus
\be
\delta_{PQ}f_{E_8}
\ap \delta_{PQ}f_{E^{\prime}_8}\ap e^{-2\pi T},
\ee
where $\delta_{PQ}$
means $U(1)_{PQ}$-breaking other than the QCD anomaly.
Let
$$
W=W_M(S,T)+\sum \frac{1}{M_P^n}\lambda_n(S,T)\Phi^{3+n}
$$
denote the superpotential
at the compactification scale where 
$\Phi$ represents   generic chiral matter multiplets (including
those in the hidden sector)
with the Yukawa-type couplings $\lambda_n(S,T)$. 
If $W_M=0$ at string tree level, which 
is the case in many interesting models including $(2,2)$ Calabi-Yau
and orbifold compactifications, it remains to be zero  
at any finite order in string perturbation theory.
(For  $(2,0)$ Calabi-Yau compactifications,
a nonzero $W_M$ may be induced by world-sheet instantons
even at string tree level \cite{dine}.) 
However non-perturbative stringy effects may generate
a nonzero $W_M\ap M_P^3 e^{-2\pi (S-lT/2)}$, and then 
$\delta_{PQ}W_M\ap M_P^3 e^{-2\pi [T+(S-lT/2)]}$.
It is known that world-sheet instantons can induce $U(1)_{PQ}$-breaking
Yukawa-type couplings at string tree level without any suppression
by $e^{-2\pi S}$ \cite{dine}.
Summarizing these, if $W_M=0$ 
at string tree level in the weak coupling limit, which is the
case that we focus on here,
the following order of magnitude estimate 
applies for $U(1)_{PQ}$-breaking
in the superpotential:
\bea
&& \delta_{PQ}W_M\ap M_P^3 e^{-2\pi [T+(S-lT/2)]}\ap
e^{-2\pi T}W_M,
\nonumber \\
&& \delta_{PQ}\lambda_n\ap e^{-2\pi T}.
\eea

The discussion of $U(1)_{PQ}$-breaking in the K\"{a}hler potential
is more subtle because of the absence of  holomorphy.
Again together with the property that
nonderivative couplings of ${\rm Im}(S)$
appear always through the combinations ${\rm Im}(f_a)
\propto {\rm Im}(S\pm \frac{1}{2}lT)$, the discrete symmetries
$Z_{S,T}$ imply that 
generic K\"{a}hler potential can be written as
$K=\sum_n K_n \exp [i2\pi n {\rm Im}(T)]$
where $K_n$ is a function of
${\rm Re}(S)$, ${\rm Re}(T)$, ${\rm Im}(S-\frac{1}{2}lT)$, and also of other
$U(1)_{PQ}$-invariant field variables.
Obviously $U(1)_{PQ}$ is broken only by $K_n$ with $n\neq 0$.
To estimate its size,
one may consider the 
limit of weakly coupled heterotic string which preserves
four-dimensional $N=2$ supersymmetry
or   the $M$-theory limit in which
the eleventh radius $R_{11}\gg R_6$ so that physics below
the energy scale $E\ler R^{-1}_6$ can be
described by a five-dimensional supergravity.
In these limits, $T$ corresponds to the coordinate
of a special  K\"{a}hler manifold whose 
K\"{a}hler potential is determined by a holomorphic prepotential
${\cal F}$ as
$K=-\ln [2({\cal F}+{\cal F}^*)-(\phi_i+\phi_i^*)(\partial_i{\cal F}+
\partial_i{\cal F}^*)]$ \cite{dewit},
and as a result
$K_n\propto e^{-2\pi nT}$ 
as  in the case of the holomorphic gauge kinetic functions
and superpotential.
As long as $Z_{S,T}$ are not spontaneously broken,
turning on  
the spontaneous breaking of four-dimensional $N=2$ supersymmetry
or of five-dimensional supersymmetry
down to four-dimensional $N=1$ supersymmetry
does not affect  this behavior of
$U(1)_{PQ}$-breaking in the large radius limit, and thus
\be
\delta_{PQ} K\ap M_P^2 e^{-2\pi T},
\ee
in generic large radius  compactifications  preserving
four-dimensional $N=1$ supersymmetry.

In the above, we have noted that supersymmetry
and the discrete gauge symmetries $Z_{S,T}$ associated with the periodicity
of the axion-like fields ${\rm Im}(S)$ and ${\rm Im}(T)$
imply that $U(1)_{PQ}$-breaking terms (other 
than the QCD anomaly) are  suppressed by $e^{-2\pi T}$
in the large radius limit.
Although it is quite reasonable to assume that
the discrete symmetries $Z_{S,T}$
are {\it not} spontaneously broken by the 
$M$-theory dynamics above the compactification scale, 
they may be broken
by  infrared
dynamics  at scales below the compactification scale.
This does  indeed occur
if non-perturbative hidden sector
dynamics  leads to the formation of
the  gaugino and/or matter condensations.

Integrating out the hidden sector gauge and matter multiplets leads
to an effective supergravity model of the visible
sector fields and also generic moduli including $S$ and $T$.
This effective supergravity will be described by the effective K\"{a}hler
potential $K_{\rm eff}$, the effective visible sector gauge kinetic
function $f_{\rm eff}$, and finally the effective superpotential
$W_{\rm eff}$ which can be written as
\be
W_{\rm eff}=W_0(S,T)+\mu (S,T)\Phi^2+
\sum \frac{1}{M_P^n}h_n(S,T)\Phi^{n+3},
\ee
where now $\Phi$ stands for  the visible sector matter multiplets.
Even for $K_{\rm eff}$, $f_{\rm eff}$, and $W_{\rm eff}$
including  nonperturbative corrections due to the field-theoretic
hidden sector dynamics,
it is rather obvious that
\be
\delta K_{\rm eff}\ap M_P^2 e^{-2\pi T}, \quad
{\delta}_{PQ}f_{\rm eff}\ap e^{-2\pi T}, \quad
\quad \delta_{PQ}h_n\ap e^{-2\pi T}. 
\ee
However one needs a further discussion to estimate 
$\delta_{PQ}W_0$ and $\delta_{PQ}\mu$. 
To proceed, let us split $W_0$ into the two
pieces as 
\be
W_0=W_M+W_F,
\ee
where $W_M$  is the piece generated by $Z_{S,T}$-preserving
$M$-theory dynamics at scales above the compactification
scale, while $W_F$ (and also $\mu$ in (32)) 
is the piece from field-theoretic infrared
effects, i.e. the gaugino and/or matter condensations, which break
$Z_{S,T}$ spontaneously.
Generically  $W_F$ and $\mu$ 
are  holomorphic functions of the hidden sector
Yukawa couplings $\lambda_n$ and the dynamical
scale
$\Lambda_{E_8^{\prime}}=M_{GUT}
\exp [-8\pi^2 f_{E_8^{\prime}}/b]$ of the hidden sector
gauge interaction.
(Here $b$ is the coefficient of the one-loop beta
function and $M_{GUT}$ can be identified as $2\pi R_6^{-1}$ in (6).)
More explicitly,
\bea
&&W_F\ap\kappa_1 M_{GUT}^3(\Lambda_{E_8^{\prime}}/M_{GUT})^{n_1/n_2}
\ap M_{GUT}^3 \exp [-8\pi^2 n_1 f_{E_8^{\prime}}/n_2 b], \nonumber \\
&&\mu\ap\kappa_2 M_{GUT} (\Lambda_{E_8^{\prime}}/M_{GUT})^{n_3/n_4}
\ap M_{GUT} \exp [-8\pi^2 n_3 f_{E_8^{\prime}}/n_4 b],
\eea
where $\kappa_1$ and $\kappa_2$ are dimensionless functions
of $\lambda_n$,
and  $n_i$'s are model-dependent
positive integers.
If there is no hidden matter  or if the hidden matter multiplets have
renormalizable Yukawa couplings, $W_F\ap 
\Lambda_{E_8^{\prime}}^3$ and thus $n_1/n_2=3$.
However if the hidden matters have only
non-renormalizable Yukawa-type couplings, $n_1/n_2$ 
can take a different value.
Note that if $n_1/n_2b$ or $n_3/n_4b$ is not integral, which is usually
the case,
$Z_S: S\rightarrow S+i$ is spontaneously
broken with the multi-valued superpotential:
$W_F\propto e^{-2\pi n_1S/n_2b}$ and $\mu
\propto e^{-2\pi n_3S/n_4b}$.
At any rate, (29) and (30) now imply 
\bea
&&\delta_{PQ}W_F=\frac{\partial W_F}{\partial f_{E^{\prime}_8}}
\delta_{PQ}f_{E^{\prime}_8}+\frac{\partial W_F}{\partial \lambda_n}
\delta_{PQ}\lambda_n\ap W_F e^{-2\pi T},
\nonumber \\
&&\delta_{PQ}\mu=
\frac{\partial \mu}{\partial f_{E^{\prime}_8}}
\delta_{PQ}f_{E^{\prime}_8}+
\frac{\partial \mu}{\partial \lambda_n}\delta_{PQ}\lambda_n
\ap \mu e^{-2\pi T}.
\eea

The axion potential due to $U(1)_{PQ}$-breaking other than
the QCD anomaly can be schematically written as
$$
\delta V_{\rm axion}= \frac{\delta V_{\rm axion}}{
\delta K_{\rm eff}}\delta_{PQ}K_{\rm eff}+
\frac{ \delta V_{\rm axion}}{\delta W_{\rm eff}}\delta_{PQ}W_{\rm eff}
+\frac{\delta V_{\rm axion}}{\delta f_{\rm eff}}
{\delta}_{PQ}f_{\rm eff}.
$$
Carefully inspecting the supergravity 
potential (18)
and also all possible 
quantum corrections including the quadratically
divergent one-loop potential
$V_{\rm loop}=\frac{1}{16\pi^2}{\rm Str}(M^2)\Lambda^2$
with the cutoff $\Lambda\ap M_{GUT}$ or $M_P$,
it is not difficult to see that
$$
\frac{\delta V_{\rm axion}}{
\delta K_{\rm eff}}
\ap m_{3/2}^2,
\quad
\frac{ \delta V_{\rm axion}}{\delta W_{\rm eff}}
\ap m_{3/2}, \quad
\frac{\delta V_{\rm axion}}{\delta f_{\rm eff}}
\ap \frac{1}{16\pi^2}m_{3/2}^2\Lambda^2,
$$
where  the main contributions to 
$\delta V_{\rm axion}/
\delta K_{\rm eff}$ and
$\delta V_{\rm axion}/\delta W_{\rm eff}$
are from the tree level supergravity potential (18) with the relation
$W_{\rm eff}\ap W_0\ap m_{3/2}M_P^2$, while  
$\delta V_{\rm axion}/\delta f_{\rm eff}$ is mainly from
$V_{\rm loop}$ through the gaugino masses which depends
upon the gauge coupling, i.e.  upon ${\rm Re}(f_{\rm eff})$.
(Here we assume that supersymmetry breaking is mainly
due to the $F$-terms, not by the $D$-terms.)
Note that $U(1)_{PQ}$-breaking in $W_{\rm eff}$
is dominated by $\delta_{PQ}W_0=\delta_{PQ}(W_M+W_F)$
which  is of order $e^{-2\pi T}W_0\ap
e^{-2\pi T}m_{3/2}M_P^2$ for $\delta_{PQ}W_M$ and $\delta_{PQ}
W_F$ estimated in (30) and (36), respectively.
Putting these together with the previously made
estimates of $U(1)_{PQ}$-breaking,
what we find is a rather simple
result:
\be
\delta V_{\rm axion} \ap 
m_{3/2}^2 M_P^2 e^{-2\pi T}.
\ee
With this, we  conclude that
if the compactification radius is 
large enough to yield
\be
{\rm Re}(T)\ger 20+\frac{1}{\pi}\ln (m_{3/2}/{\rm TeV}),
\ee
$U(1)_{PQ}$ breaking 
other than the QCD anomaly,
{\it whatever its microscopic origin is},
is suppressed enough to satisfy the condition
(1) for the strong CP problem
to be solved by the axion mechanism.

As we have noted, it is more precise 
to interprete  our estimates of $U(1)_{PQ}$-breakings  
as an approximate upper limit.
One might then  interprete 
the estimate of the high energy
axion potential in (37)  also as an upper limit.   
However as long as 
anyone of $\delta_{PQ}K$,  
$\delta_{PQ}f_{E_8^{\prime}}$,
and $\delta_{PQ}\lambda$
saturates their estimated  upper limits,
which is true for the most of known compactified
string models,
(37) corresponds to a correct order of magnitude
estimate, not merely an upper limit.
(Here $\lambda$ corresponds to the lowest dimensional
non-vanishing Yukawa coupling of hidden matter.)
If only $\delta_{PQ}f_{E_8}$ or $\delta_{PQ} h_0$ saturates
the bound, the resulting high energy axion potential
would be  of order of $m_{3/2}^2\Lambda^2/(16\pi^2)^k$ where
$k=1$ for $f_{E_8}$ and $k=2$ for the visible matter
Yukawa coupling $h_0$.

Our argument  in this section  can be easily generalized to  
the  cases with a semisimple hidden sector
gauge group and/or  the number of the
model-dependent K\"{a}hler axions  $h_{1,1}> 1$.
Note that one may need a semisimple hidden sector 
gauge group  with nontrivial matter
contents in order to stabilize the dilaton and moduli vacuum
expectation values through 
the racetrack mechanism
\cite{race}. 
For the gauge kinetic functions written as (7),
one can define a  $U(1)_{PQ}$-symmetry similarly as (17),
e.g.
$$
U(1)_{PQ}: \quad S\rightarrow S+i\alpha,
\quad T_I\rightarrow T_I+ik_I\alpha,
$$
where the real coefficients $k_I$ are chosen to make this $U(1)_{PQ}$
to be free from any of the hidden sector anomalies, while  be broken
by the QCD anomaly.
Such a $U(1)_{PQ}$ exists always if $h_{1,1}\geq N_H$
where $N_H$ denotes the number of simple 
gauge groups in hidden sector. It would exist
even when $h_{1,1}< N_H$ if some of the hidden sector
gauge kinetic functions are not linearly-independent from each other.
Then  $U(1)_{PQ}$-breaking other than the QCD anomaly will be 
suppressed enough if the compactification radius
is large enough so that  all $T_I$ with 
$\delta_{PQ}T_I\neq 0$  satisfy 
the condition (38).

\section{Phenomenological Constraints and Axion Cosmology}

In the previous section, we have argued  
that  if the compactification
radius is large enough so that  
\be
{\rm Re}(T_I)\ger 20+\frac{1}{\pi}\ln (m_{3/2}/{\rm TeV})
\ee
for all $T_I$ with $\delta_{PQ}T_I\neq 0$,
the strong CP problem can be  solved by  string-theoretic
axions.
Perhaps the most serious difficulty
with this large radius compactification 
would be to stabilize the dilaton and moduli  at the 
desired vacuum values.
Here we simply assume that
the dilaton and moduli can be stabilized at a point
satisfying (39),
and just look at its phenomenological viability.

For the gauge kinetic functions written as (7),
most of the known heterotic string models (and  thus
their $M$-theory limits also)  give $k_a=1$ \cite{dines}.
Let  $f_{E_8}$  of Eq. (11)  denote the  QCD gauge
kinetic function and 
$f_{E_8^{\prime}}$ denote the gauge kinetic
function for the hidden sector gauge interaction 
which gives a
dominant contribution to the field-theoretic 
nonperturbative superpotential $W_F$ in (34).
Then the
phenomenological value  of $\alpha_{QCD}$ at $M_{GUT}$  gives
\be
4\pi {\rm Re}(f_{E_8})\ap 25. 
\ee
We also find
$$
4\pi {\rm Re}(f_{E^{\prime}_8})\ap \frac{n_2b}{2\pi n_1}
\ln (M_{GUT}^3/|W_F|)\ap \frac{n_2b}{n_1}[4.4
+\frac{1}{2\pi}\ln({\rm TeV}/m_{3/2})],
$$
using $M_{GUT}\ap 2\pi R_6^{-1}\ap 2\times 10^{17}$ GeV, 
and also  the expression of $W_F$ in (35) 
together with the assumption that
supersymmetry breaking  is mainly
due to $W_F$,
not due to the
$M$-theoretic non-perturbative term $W_M$ in (34),
and thus $W_F\ap m_{3/2}M_P^2$.
(This assumption is not so crucial
for our discussion.)
Combining these with (11),
we  obtain
\be
\frac{1}{2}\sum (l_I-l^{\prime}_I){\rm Re}(T_I)\ap
25- \frac{n_2b}{n_1}[4.4
+\frac{1}{2\pi}\ln({\rm TeV}/m_{3/2})]
 -{\rm Re}(\triangle_{E_8}-
\triangle_{E^{\prime}_8}),
\ee
where $\frac{1}{2}(l_I-l^{\prime}_I)$ are integral as
required by  (10).

In order to have a $U(1)_{PQ}$-symmetry which avoids the hidden sector
anomaly, one needs at least
one of $\frac{1}{2}(l_I-l^{\prime}_I)$ to be a nonzero integer.
Furthermore, as we have noticed,
$\triangle_a$ in the gauge kinetic function
corresponds to higher order correction in
the strong coupling expansion and thus ${\rm Re}(\triangle_a)$
is essentially of order one or less in the $M$-theory limit
with ${\rm Re}(T_I)\gg 1$ \cite{kaplu}.
Let us also recall that $b$ is the positive coefficient
of  one-loop beta function, and $n_1$ and $n_2$
are model-dependent positive integers.
Then comparing 
the large radius condition (39) 
with the phenomenological relation (41), we easily
find that they can be compatible only for
a rather limited set of the coefficients  
$\{\frac{1}{2}(l_I-l^{\prime}_I)\}$.
For instance, if anyone of $\frac{1}{2}(l_I-l^{\prime}_I)$ is significantly
bigger than one, there has to be another
coefficient  with a 
similar magnitude but with a different sign.

In the above, we have noted that
the phenomenological viability of the large radius condition (39)
crucially depends  upon
the quantized coefficients of
the K\"{a}hler moduli superfields $T_I$ in
gauge kinetic functions,
and in fact it is viable 
only for a rather restricted class of models.
This is particularly true for
supersymmetry-preserving compactifications 
on a smooth Calabi-Yau manifold
with vanishing $E^{\prime}_8$ field strength.
The hidden sector of such compactifications
does not contain any matter multiplet,
and then $n_2b/n_1$  corresponds to
the second Casimir $C_2={\rm tr}(T^2_{\rm adj})$ of  the gauge group
giving a dominant contribution to the field-theoretic
nonperturbative superpotential $W_F$ through
the gaugino condensation.
Since the hidden sector gauge group $G_h$ is
a subgroup of $E^{\prime}_8$ commuting  
with the Wilson lines in the model,
we have $2\leq C_2\leq 30$.
(If  $G_h=\prod SU(N_i)\times
\prod U(1)$ for instance, $C_2={\rm Max}(N_i)$.)
In sect. III, we have shown that $l_I=-l^{\prime}_I$ in
generic compactifications on a smooth six-manifold.
It is also noted that
if $F^{\prime}_{AB}=0$,
then one of the coefficients $\frac{1}{2}(l_I-l^{\prime}_I)=l_I$ 
is a positive integer,
while the other coefficients  are still non-negative
(see Eq. (15)).
Summarizing these, for supersymmetry-preserving compactifications
on a smooth Calabi-Yau manifold with $F^{\prime}_{AB}=0$, we have
\be
\sum l_I{\rm Re}(T_I)
\ap 
25-C_2[4.4+\frac{1}{2\pi}\ln({\rm
TeV}/m_{3/2})]
-{\rm Re}(\triangle_{E_8}-
\triangle_{E^{\prime}_8}),
\ee
where at least one of the non-negative integers
$l_I=\int \omega^I\wedge I_4$
is non-zero,
and $2\leq C_2\leq 30$.
Since ${\rm Re}(\triangle_a)$ corresponds to a subleading
part of order one or less,
the above result 
implies that  (i) the nonvanishing $l_I$ has to be fixed to be one,
and (ii) $C_2$ can {\it not} be significantly bigger than
its minimal value two, in order for the large radius condition
(39) to be satisfied.
In view of the boundary condition $\frac{1}{\pi}[G]_{E_8}=I_4$
(see the discussion below Eq. (14)),
this is possible only when  (i) the quantized flux
of the antisymmetric tensor field in $M$-theory
has the minimal nonzero value: $[G/2\pi]=1/2$
in the notation of ref. \cite{witten3}, and
(ii) the hidden gauge group $E^{\prime}_8$ is broken
by Wilson lines to a  subgroup with small values of the second  
Casimir $C_2={\rm tr}(T^2_{\rm adj})$.

As is well known, the QCD axion with a decay constant
$v\gg  10^{12}$ GeV
can be cosmologically troublesome \cite{kim}.
Let us assume that axionic strings were inflated away
in the early stage, and thus ignore the relic axions
emitted from axionic strings.
However still the coherent axion oscillation after the QCD
phase transition
in the early universe gives rise to relic axions
which may overclose
the universe at the present \cite{pww}.
If there is no entropy production after the QCD phase transition,
the relic axion mass density (in the unit of
the critical energy density) at the present
is given by \cite{tur}
\be
\Omega_a \ap
\left(\frac{\delta \theta}{3\times 10^{-3}}\right)^2
\left(\frac{v}{10^{16} \, {\rm GeV}}\right)^{1.18}\left(
\frac{\Lambda_{QCD}}{200 \, {\rm MeV}}\right)^{-0.7},
\ee 
where $\delta\theta=\delta a/v$ denotes the misalignment
angle of the axion field at the time of QCD phase transition
in the early universe.
For the case with a late-time entropy production, the  relic axions
are diluted as \cite{lps}
\be
\Omega_a \ap
\delta\theta^2
\left(\frac{v}{10^{16} \, {\rm GeV}}\right)^{1.5}
\left(\frac{T_{RH}}{6 \, {\rm MeV}}\right)^2
\left(\frac{\Lambda_{QCD}}{200 \, {\rm MeV}}\right)^{-2},
\ee
where the big-bang nucleosynthesis requires 
the reheat temperature $T_{RH}\ger 6$ MeV.

The above formulae for the relic axion energy density
indicates that if $v\gg  10^{12}$ GeV,
one needs either a mechanism for 
the axion misalignment $\delta\theta\ll 1$, or
a late time entropy production, or both.
They also imply that $v\gg 10^{16}$ GeV
should be distinguished  from $v\ler 10^{16}$ GeV.
In order to be cosmologically viable, the former requires
a significant suppression of $\delta\theta$ independently of
whether there is a late time entropy production or not,
while the latter can be viable only with a
late time entropy production with $T_{RH}\ger 6$ MeV.

The late time entropy production has been suggested
a long time ago as a mechanism to make $v\gg 10^{12}$ GeV
cosmologically viable \cite{stein}.
In fact, it can occur naturally in string effective
supergravity models \cite{lps,thermal}.
For instance, 
moduli with $m=O(10)$ TeV or the flaton fields triggering thermal inflation
lead to a huge entropy production after 
the QCD phase transition but still before the 
big-bang nucleosynthesis 
\cite{bank3,choi2}.
It was  argued also
that $\delta\theta$ may be relaxed
down to a small value if there is a period in the early universe
during which the expectation values
of some moduli fields differ
from the present ones and as a consequence
a large effective axion mass $m_{\rm eff}$ is induced \cite{dvali,bank3}.
If $m_{\rm eff}$ is bigger than the Hubble expansion rate
$H$, the axion field would be driven
to the vacuum value in this period.
However usually the moduli values to raise up the axion mass raise
up also the vacuum energy density \cite{choi3}, and thereby it is difficult
to arrange $m_{\rm eff}\ger H$.
Furthermore, the axion vacuum value in this period  generically
differs from the present one by the order of $v$,
which would result in $\delta\theta\ap 1$,  {\it unless}
the expectation values of the moduli which  affect CP violating phases
are (approximately) same  as the present values.
Due to these difficulties, the mechanism for $\delta\theta\ll 1$
appears to involve  too many cosmological assumptions.

It is rather obvious that $v\gg 10^{12}$ GeV for 
the QCD axion in $M$-theory limit.
However in view of the above discussion,
it is still a relevant question to ask
whether $v\gg 10^{16}$ GeV or $v\ler 10^{16}$ GeV.
To answer this question, let us
estimate  the QCD axion decay constant in a simple model
with $h_{1,1}=1$.
It has been pointed out recently that a simple dimensional
reduction of eleven-dimensional supergravity leads to the
K\"{a}hler potential of $S$ and $T$ 
which is very similar to the one obtained 
in  weakly coupled heterotic string theory  
\cite{nano}:
$$
K=-\ln (S+S^*)-3\ln (T+T^*).
$$
This  K\"{a}hler potential may  receive a sizable  
$M$-theoretical correction, however it is expected that
our follwing analysis is not significantly affected by
this correction as long as ${\rm Re}(S)\gg 1$ and ${\rm Re}(T)\gg 1$.  
Using the above K\"{a}hler potential and the gauge kinetic functions
in (26) with $l=1$, we find 
$$
{\cal L}_{\rm axion}= \frac{1}{2}(\partial_{\mu}a)^2
+\frac{1}{2}(\partial_{\mu}a^{\prime})^2+
\frac{1}{32\pi^2}\frac{a}{v}
F^{\mu\nu}\tilde{F}_{\mu\nu}
+\frac{1}{32\pi^2}\frac{a^{\prime}}{v^{\prime}}
F^{\prime\mu\nu}\tilde{F}^{\prime}_{\mu\nu}+...,
$$
where the ellipsis denotes the irrelevant
terms including the high energy potential $\delta V_{\rm axion}$
which has been argued to be of order
$e^{-2\pi T}m_{3/2}^2 M_P^2$ in the previous section, 
and the axion fields
$a$ and $a^{\prime}$ are defined as 
$$
a=\frac{2\pi(v_{_S}^2{\rm Im}(S)+v_{_T}^2{\rm Im}(T)/2)}{\sqrt{v^2_{_S}
+v_{_T}^2}},
\quad
a^{\prime}=\frac{2\pi v_{_S}v_{_T}({\rm Im}(S)-{\rm Im}(T)/2)}{\sqrt{v_{_S}^2
+v_{_T}^2}},
$$
for the axion scales 
\be
v_{_S}=\frac{M_P}{2\pi \sqrt{2}\langle{\rm Re}(S)\rangle},
\quad
v_{_T}=\frac{\sqrt{3}M_P}{\pi\sqrt{2}\langle{\rm Re}(T)\rangle}.
\ee
Here the axion decay constants $v$ and $v^{\prime}$ are given by  
\be
v=\frac{1}{2}\sqrt{v_{_S}^2+v_{_T}^2},
\quad
v^{\prime}=\frac{v_{_S}v_{_T}}{\sqrt{v_{_S}^2+v_{_T}^2}}.
\ee

The field combination $a$  corresponds
to the QCD axion which would solve the strong CP
problem if ${\rm Re}(T)$ satisfies the large radius condition (39),
while the other combination $a^{\prime}$ couples to the hidden
sector anomaly and thus irrelvant for the strong CP problem.
Here we assumed that the axion potential 
due to the hidden sector anomaly is much bigger than the 
QCD induced potential
$V_{QCD}\ap f_{\pi}^2 m_{\pi}^2$, 
and thus $m_{a^{\prime}}\gg m_a$.
Note that in models with multi-gaugino
(matter) condensations which are comparable to each other,
the axion potential due to the hidden sector
anomaly is of
order $m_{3/2}^2M_P^2$, and then
the hidden sector axion $a^{\prime}$ receives  
a mass of order $m_{3/2}M_P/v^{\prime}
\ap 10^2 m_{3/2}$.

In $M$-theory limit, we have roughly 
${\rm Re}(S)\ap {\rm Re}(T)\ap 1/\alpha_{GUT}$.
Let us recall that the phenomenological relations (40) and (41) suggest 
that neither ${\rm Re}(S)$ nor ${\rm Re}(T)$ can be significantly
bigger than $1/\alpha_{GUT}$.
We then find from (45) and (46) that the
QCD axion decay constant in $M$-theory limit is given by
\be
v\ap 5\times 10^{15} \sqrt{(25/{\rm Re}(S))^2+
(85/{\rm Re}(T))^2} \, {\rm GeV}
\ap 2\times 10^{16} \, GeV.
\ee
Here we stress that  
the large radius
with ${\rm Re}(T)\ap 1/\alpha_{GUT}$ is crucial for 
$v\ap 10^{16}$ GeV, {\it not} the order of $M_P$, so that 
the QCD axion can be cosmologically viable
in the presence of
a late time entropy production without assuming any significant
suppression of the axion misalignment angle. 
If ${\rm Re}(T)\ll 1/\alpha_{GUT}$, the axion scale $v_{_T}$
of the model-dependent K\"{a}hler axion would be of order
$M_P$ 
although the axion scale $v_{_S}$ of the model-independent
axion is still smaller than $M_P$ by two orders
of magnitudes: $v_{_S} \ap \frac{\alpha_{GUT}}{4\pi}M_P\ap 10^{16}$ GeV.
In this case, the QCD axion decay constant $v$ is essentially given by
$v_{_T}\gg v_{_S}$, and as a result
it would be of order $M_P$ if 
${\rm Re}(T)\ler 1$ for instance.

The energy density crisis of the QCD axion in $M$-theory
may be ameliorated if there is an {\it accidental}
axion-like field with a smaller decay
constant \cite{choi}.
Let  $a_1$ denote the string-theoretic QCD axion  which 
corresponds to a linear combination
of the model-independent axion ${\rm Im}(S)$ and the
model-dependent K\"{a}hler axions ${\rm Im}(T_I)$.
As discussed above, 
the decay constant and the high energy potential of $a_1$ obey 
$v_1\ap 10^{16}$ GeV and
$\delta V_{a_1}\ap e^{-2\pi T}m_{3/2}^2M_P^2\ler 10^{-9}
f_{\pi}^2m_{\pi}^2$ for 
${\rm Re}(T)\ap 1/\alpha_{GUT}$.
Let us suppose that 
there is an {\it accidental} axion-like field 
$a_2$ with an unspecified high energy potential
$\delta V_{a_2}$ and the decay constant $v_2\ll v_1$.
The axion effective lagrangian is then given by:
$$
{\cal L}_{\rm axion}=\frac{1}{2}
(\partial_{\mu}a_1)^2+\frac{1}{2}(\partial_{\mu}a_2)^2
+\frac{1}{32\pi^2}(\frac{a_1}{v_1}+\frac{a_2}{v_2})
F^{\mu\nu}\tilde{F}_{\mu\nu}
-\delta V_{a_2},
$$
where we have ignored the high energy potential of $a_1$.

In this case, we have  two pseudo-Goldstone bosons 
whose   relic mass densities would be determined by the mass
eigenvalues 
$m_i$ and
the corresponding axion misalignments $\delta a_i$ ($i=1,2$).
Let us examine how the presence of the accidental axion $a_2$
affects the cosmology of the string-theoretic QCD  axion $a_1$.
If $\delta V_{a_2}\ger f_{\pi}^2m_{\pi}^2$, 
we have 
$$
(m_1\ap \frac{f_{\pi}m_{\pi}}{v_1},\quad  \delta a_1\ap v_1),
\quad (m_2\ap \frac{\sqrt{\delta V_{a_2}}}{v_2},\quad  \delta a_2\ap v_2).
$$
Obviously then  having  
$a_2$ with $v_2\ll v_1$  changes neither the
mass nor the initial misalignment of the original axion $a_1$,
and as a result our previous discussion of the relic mass density
of the string-theoretic QCD axion remains to be valid.
Furthermore, in this case the accidental axion-like field $a_2$
can lead to its own   cosmological problem {\it unless}
$m_2$ is large enough (or $v_2$ is small enough).

In the case with $\delta V_{a_2}\ll f_{\pi}^2 m_{\pi}^2$,
the mass eigenvalues  and the  misalignments are given by
$$
(m_1\ap \frac{\sqrt{\delta V_{a_2}}}{v_1}, \quad \delta a_1\ap v_1),
\quad (m_2\ap \frac{f_{\pi}m_{\pi}}{v_2}, \quad  \delta a_2\ap v_2).
$$
Then $a_2$ corresponds to the true QCD axion with a decay constant
$v_2\ll v_1\ap 10^{16}$ GeV,
which would be cosmologically safe without any late time 
entropy production or any suppression of the misalignment
angle $\delta\theta_2=\delta a_2/v_2$ if $v_2\ler 10^{12}$ GeV.
However the other pseudo-Goldstone boson which is mainly
the string-theoretic axion $a_1$   can still be cosmologically
troublesome. If there is no entropy production below
the temperature $T_{\rm osc}\ap\sqrt{m_1M_P}$, the relic mass density
of $a_1$ (again in the unit of the critical density) at the present
would be given by
$$
\Omega_{a_1}\ap 5\times 10^3 
\left(\frac{\delta V_{a_2}}{f_{\pi}^2m_{\pi}^2}\right)^{1/4}\left(
\frac{v_1}{10^{16} {\rm GeV}}\right)^{1.5}
\left(\frac{\delta a_1}{v_1}\right)^2.
$$ 
Although there is a  suppression of the relic mass density by  the  
small factor
$(\delta V_{a_2}/f^2_{\pi}m^2_{\pi})^{1/4}$,
this  would not be so significant 
{\it unless} $\delta V_{a_2}/f_{\pi}^2 m_{\pi}^2$  is extremely small.
For instance, in the absence of a late time 
entropy production after $T_{\rm osc}\ap
\sqrt{m_1M_P}$ and also of  any suppression of the misalignment
angle $\delta\theta_1=\delta a_1/v_1$, the pseudo-Goldstone
boson $a_1$ would be cosmologically safe only when
$\delta V_{a_2}\ler 10^{-15}f_{\pi}m_{\pi}^2$.
In view of the absence of an exact continuous global symmetry in string
theory \cite{global}, 
such an extremely small  high energy potential 
of the accidental axion-like field 
is highly unlikely.
We thus conclude that having an accidental axion like field with a
decay constant much smaller than $10^{16}$ is not so helpful
for ameliorating the cosmological difficulty
of the QCD axion in $M$-theory whose decay constant
was estimated to be of order $10^{16}$ GeV.

{\bf Acknowledgments}:
This work is supported in part by 
KAIST Basic Science Research
Program, KAIST Center for Theoretical Physics and Chemistry, 
KOSEF through CTP of
Seoul National University, Distinguished 
Scholar  Exchange Program of KRF, and  Basic Science 
Research Institutes Program BSRI-97-2434.

\end{document}